\documentstyle[12pt]{article}

\newcommand{\be}{\begin{equation}}
\newcommand{\ee}{\end{equation}}
\newcommand{\ba}{\begin{eqnarray}}
\newcommand{\ea}{\end{eqnarray}}
\newcommand{\bann}{\begin{eqnarray*}}
\newcommand{\eann}{\end{eqnarray*}}

\renewcommand{\hfill}{\hspace*{\fill}}
\begin{document}
\hbadness=10000
\thispagestyle{empty}
\begin{samepage}
\title{
\hfill {\normalsize DMR-THEP-93-5/W}\\*[-1.5ex]
\hfill {\normalsize hep-ph/9311230 }\\*[-1.5ex]
\hfill {\normalsize October 1993} \\[2ex]
Bounds for Bose-Einstein correlation functions }
\author{
M. Pl\"umer\thanks{E.Mail: pluemer\_m@vax.hrz.Uni-Marburg.de},\ \ \
L.V. Razumov\thanks{E.Mail: razumov@convex.hrz.Uni-Marburg.de},\ \ \
R.M. Weiner\thanks{E. Mail: weiner@vax.hrz.Uni-Marburg.de} }

\date{Physics Department, University of Marburg, Marburg, F.R. Germany}
\maketitle
\thispagestyle{empty}
\begin{center}
{\large\bf Abstract}\\[2ex]
\end{center}
\noindent
Bounds for the correlation functions of identical bosons are discussed
for the general case of a Gaussian density matrix.  In particular, for
a purely chaotic system the two-particle correlation function must
always be greater than one.  On the other hand, in the presence of a
coherent component the correlation function may take values below
unity. The experimental situation is briefly discussed.\\


\end{samepage}
\newpage

The most important method to obtain information about space-time
aspects of particle production in high energy collisions is particle
interferometry. One studies the correlation function of $m$ ($m\geq
2$) identical particles,
\begin{equation}
C_m(\vec{k}_1,...,\vec{k}_m) \ =\ \frac {P_m(\vec{k}_1,...,\vec{k}_m)}
{P_1(\vec{k}_1) \cdot ...
\cdot P_1(\vec{k}_m)}
\end{equation}
where $\vec{k}_i$ is the three-momentum of the $i$-th particle and
$P_\ell(\vec{k}_1,...,\vec{k}_\ell)$ are the $\ell$-particle single
inclusive distributions.  For ultrarelativistic collisions, one
usually considers bosons, i.e., one measures Bose-Einstein
correlations (BEC).

In practice most of the experimental measurements have been restricted
so far to two particle correlations (see, however, ref. \cite{bias}).
Moreover, with very few exceptions, only identical charged particles
were considered.  Usually the data are analysed by fitting them to a
form
\begin{equation}
C_2(\vec{k}_1,\vec{k}_2)\ =\ 1 \ + \
\left|\tilde{\rho}(k_1-k_2)\right|^2
\label{eq:simple}
\end{equation}
where $\tilde{\rho}(q)$ is the Fourier transform of the space-time
distribution of the source elements,
\begin{equation}
\tilde{\rho}(q) \ =\ \int d^4x \ \rho(x) \ \exp(iq_\mu x^\mu)
\end{equation}
Thus, the inverse widths of the correlation function are related to
the size and lifetime of the source.  Eq. (\ref{eq:simple}) was first
derived by symmetrizing the two particle wave function and averaging
over the emission points of the particles {\it after} taking the
square of the matrix element \cite{gglp}.  This corresponds to the
assumption of a purely chaotic source (random phases).  {}From Eq.
(\ref{eq:simple}) one obtains
\begin{equation}
1 \ \leq \ C_2(\vec{k}_1,\vec{k}_2) \ \leq \ 2
\label{eq:bounds}
\end{equation}
and these are the bounds for $C_2$ in the conventional treatment of
BEC.  For neutral particles as well as for a partially coherent
source, modifications of these bounds appear (cf. below).  The
existence of bounds for BEC is important, among other things, because
they can serve as checks for results derived in any particular
approach.

A more general approach to BEC is based on the density matrix
formalism, i.e., a formalism which does not assume pure states (wave
functions).
To be specific, we shall consider the case (which covers almost all
phenomena known in quantum statistics) that the multiparticle system
is described by a {\it Gaussian density matrix}
\footnote{We do not treat final state interactions here.}.
If the particles are emitted {}from a large number of independent
source elements, such a form of the density matrix follows {}from the
central limit theorem (cf., e.g., \cite{saleh}).  Upper bounds for the
two-particle correlation function for a Gaussian density matrix have
already been discussed in \cite{apw}: for charged bosons, one finds
again $C_2(\vec{k}_1,\vec{k}_2) \leq 2$, but for some neutral bosons
like $\pi^0$'s and photons, $C_2(\vec{k}_1,\vec{k}_2) \leq 3$.  As
will be demonstrated below, concerning the lower bounds one can derive
the following general results: (a) for a purely chaotic source,
$C_2(\vec{k}_1,\vec{k}_2) \geq 1$, and (b) for a partially coherent
source, $C_2(\vec{k}_1,\vec{k}_2) \geq 2/3$ for identical charged
bosons, while for $\pi^0$'s and photons, $C_2(\vec{k}_1,\vec{k}_2)
\geq 1/3$.

For a Gaussian density matrix, all multiparticle inclusive
distributions can be expressed \cite{apw} in terms of the quantities
\begin{eqnarray}
D(k_r,k_s) & \equiv & \sqrt{E_rE_s} \
\langle a^\dagger (\vec{k}_r) a (\vec{k}_s)  \rangle \\
\tilde{D}(k_r,k_s) & \equiv & - \sqrt{E_rE_s} \
\langle a (\vec{k}_r) a (\vec{k}_s)  \rangle \\
I(k_r) & \equiv & -i \sqrt{E_r} \
\langle a (\vec{k}_r) \rangle
\end{eqnarray}
where $a^\dagger (\vec{k})$ and $a(\vec{k})$ are the creation and
annihilation operators of a particle of momentum $\vec{k}$, and the
indices $r,s$ label the particles.

For the general case of a partially coherent source, the single
inclusive distribution
can be expressed as the sum of a chaotic component and a coherent
component,
\begin{equation}
P_1(\vec{k}) \ = \ P_1^{chao}(\vec{k})\ +\ P_1^{coh}(\vec{k})
\label{eq:sig1}
\end{equation}
with
\begin{equation}
P_1^{chao}(\vec{k})
\ =\ D(k,k)
\label{eq:ncha}
\end{equation}
and
\begin{equation}
P_1^{coh}(\vec{k})
\ =\  |I(k)|^2
\label{eq:nco}
\end{equation}

\noindent
The chaoticity parameter $p$, representing the ratio between the mean
number of chaotically produced particles and the mean total
multiplicity, is then
\begin{equation}
p(k) \ =\ \frac{D(k,k)}{D(k,k)+|I(k)|^2}
\label{eq:fk}
\end{equation}

To write down the correlation functions in a concise form, it is
useful to introduce the normalized current correlators,
\begin{equation}
d_{rs} = \frac{D(k_r,k_s)}{[D(k_r,k_r)\cdot
D(k_s,k_s)]^{\frac{1}{2}}},
\quad
\tilde{d}_{rs} = \frac{\tilde{D}(k_r,k_s)}{[D(k_r,k_r)\cdot
D(k_s,k_s)]^{\frac{1}{2}}},
\label{eq:drs}
\end{equation}
where the indices $r,s$ label the particles. Since $d(k,k')$ and
$\tilde{d}(k,k')$ are in in general complex valued, one may prefer to
express the correlation functions in terms of the magnitudes and the
phases,
\begin{eqnarray}
T_{rs} &\equiv & T(k_r,k_s) \ =\ |d(k_r,k_s)|\nonumber\\
\tilde{T}_{rs} &\equiv& \tilde{T}(k_r,k_s) \ = \
| \tilde{d}(k_r,k_s)| \nonumber \\
\phi^{ch}_{rs} &\equiv& \phi^{ch}(k_r,k_s)\ =\
 {\rm Arg} \ d(k_r,k_s)\\
\tilde{\phi}^{ch}_{rs} &\equiv& \tilde{\phi}^{ch}(k_r,k_s)
\ =\  {\rm Arg} \ \tilde{d}(k_r,k_s) \nonumber
\label{eq:trs}
\end{eqnarray}
and the phase of the coherent component,
\begin{equation}
\phi^c_r  \ \equiv \ \phi^c(k_r) \ =\ {\rm Arg}\  I(k_r)\nonumber \, .
\label{eq::phic}
\end{equation}
The same notation will be employed for the chaoticity parameter,
\begin{equation}
p_r \ \equiv \ p(k_r)
\label{eq:pr}
\end{equation}

For identical charged bosons (e.g., $\pi^-$) the two-particle
correlation function reads
\begin{equation}
C^{--}_2(\vec{k}_1,\vec{k}_2)\ =\ 1 +\ 2 \sqrt{p_1(1-p_1)\cdot
p_2(1-p_2)}
\ T_{12}\  \cos(\phi^{ch}_{12} - \phi^c_1+\phi^c_2)\
+\ p_1p_2 \ T^2_{12}
\label{eq:c2plus}
\end{equation}

For neutral bosons like photons, or $\pi^0$'s, the terms
$\tilde{d}(k_r,k_s)$ may contribute\footnote{These terms are expected
to play a significant role only for soft particles of energies of the
order of the inverse lifetime of the source, or for sources with very
small lifetimes \cite{apw}.} to the BEC function\cite{apw}:
\begin{eqnarray}
C^{00}_2(\vec{k}_1,\vec{k}_2) & = & 1 + 2 \sqrt{p_1(1-p_1)\cdot
p_2(1-p_2)}
\ T_{12}\  \cos(\phi^{ch}_{12} - \phi^c_1+\phi^c_2)\
+\ p_1p_2 \ T^2_{12}\nonumber\\ & & + 2 \sqrt{p_1(1-p_1)\cdot
p_2(1-p_2)}
\ \tilde{T}_{12}\  \cos(\tilde{\phi}^{ch}_{12} - \phi^c_1-\phi^c_2)
\ + \  p_1p_2 \  \tilde{T}^2_{12}\nonumber\\
&&
\label{eq:cnulnul}
\end{eqnarray}

Let us first consider the case of a purely chaotic source. Insertion
of $p(k) \equiv 1$ in Eqs. (\ref{eq:c2plus}) and (\ref{eq:cnulnul})
immediately yields $C_2(\vec{k}_1,\vec{k}_2)\geq 1$.  In the case of
partial coherence, the terms containing cosines come into play and
consequently $C_2$ may take values below unity.  Eqs.
(\ref{eq:c2plus},\ref{eq:cnulnul}) imply that
$C^{--}_2(\vec{k}_1,\vec{k}_2)\geq 2/3$ and
$C^{00}_2(\vec{k}_1,\vec{k}_2)\geq 1/3$.  Because of the cosine
functions in (\ref{eq:c2plus},\ref{eq:cnulnul}) one would expect $C_2$
as a function of the momentum difference $q$ to oscillate between
values above and below $1$. Indeed such a behaviour of the
Bose-Einstein correlation function has been observed in high energy
$e^+e^-$ collision experiments (cf., e.g., ref. \cite{osc}), but
apparently not in hadronic reactions. This observation was interpreted
as a consequence of final state interactions in ref. \cite{bowler}. If
final state interactions determine this effect, it is unclear why the
effect is not seen in hadronic reactions. On the other hand, if
coherence is responsible for it, this would be easier to understand.
Indeed multiplicity distributions of secondaries in $e^+e^-$ reactions
are much narrower (almost Poisson-like) than in $pp$ reactions, which
is consistent with the statement that hadronic reactions are more
chaotic than $e^+e^-$ reactions \cite{fried}.

So far, two methods have been proposed for the detection of coherence
in BEC: the intercept criterion \cite{gnf} ($C_2(\vec{k},\vec{k})<2$)
and the two exponent structure of $C_2$ \cite{rm}. Both these methods
have their difficulties because of statistics problems and other
competing effects. The observation of $C_2(\vec{k}_1,\vec{k}_2)<1$
could constitute a third criterion for coherence, although
$C_2(\vec{k}_1,\vec{k}_2)<1$ is not a necessary condition.

Recently, the two-particle correlation function has been calculated
for photons emitted {}from a longitudinally expanding system of hot
and dense hadronic matter created in ultra\-rela\-ti\-vist\-ic nuclear
collisions \cite{kapusta}. For such a system, the particles are
emitted {}from a large number of independent source elements (fluid
elements), and consequently, one would expect the multiparticle final
state to be described by a Gaussian density matrix.  However, although
the system is assumed to be purely chaotic (i.e., it does not contain
a coherent component) the correlation function calculated in
\cite{kapusta} is found to take values significantly below unity.
Clearly, this is in in contradiction with the general result derived
above {}from quantum statistics ($C_2\geq 1$ for a chaotic system).

The expression for the two-particle inclusive distribution used in
ref. \cite{kapusta} (equation (3) of that paper), which in our
notation takes the form ($w(x,k)$ is the emission rate)
\begin{equation}
P_2(\vec{k}_1,\vec{k}_2) \ =\ \int d^4x_1 \int d^4x_2\
w\left(x_1,k_1\right)\ w\left(x_2,k_2\right)\ [1\ +\
\cos((k_1-k_2)(x_1-x_2))]
\label{eq:3}
\end{equation}
does not exclude values below unity for the two-particle correlation
function\footnote{To see this, consider, e.g., the simple ansatz
$$w(x,k)\ =\ const. \
\exp[-\alpha(\vec{x}-\beta \vec{k})^2]\ \delta(t-t_0)
$$ where $\alpha$ and $\beta$ are free parameters.  The expression for
$P_2(\vec{k}_1,\vec{k}_2)$ used in ref.\cite{kapusta} (cf. eq.
(\ref{eq:3})) then yields $$ C_2(\vec{k}_1,\vec{k}_2) \ =\ 1 \ +\
\exp\left[-\frac{\vec{q}^{\ 2}} {2\alpha}\right] \ \cos[\beta
\vec{q}^2] $$ Clearly, if $\beta$ exceeds $\alpha^{-1}$ the above
expression will oscillate and take values below unity. On the other
hand, in the current formalism (cf. below) one obtains with the same
ansatz for $w$ $$ C_2(\vec{k}_1,\vec{k}_2) \ =\ 1 \ +\
\exp\left[-\frac{\vec{q}^{\ 2}} {2\alpha}\right] \ \geq \ 1.  $$ }.
Thus, the fact that the authors of ref. \cite{kapusta} find values
below one for $C_2$ may be due to the application of an inadequate
expression for $P_2(\vec{k}_1,\vec{k}_2)$ \footnote{Another
possibility which was pointed out by the referee of the present paper
is that the values below one obtained for $C_2$ in the calculations of
ref. \cite{kapusta} are caused by a certain approximation used to
evaluate the integrals in
\cite{kapusta}.}
\footnote{In a recent publication \cite{kapusta1} the authors of
ref. \cite{kapusta} have included the effects of transverse expansion.
The resultant correlation functions do not take values below one.
However, if the results for the 1-dimensional expansion are affected
by the use of an inadequate expression for $P_2(\vec{k}_1,\vec{k}_2)$,
this would also cast doubts on the results obtained for the
3-dimensionally expanding system.}.

Below, we shall briefly derive expressions for
$P_2(\vec{k}_1,\vec{k}_2)$ which could be used for the problem treated
in \cite{kapusta,kapusta1} and which do not suffer {}from the
deficiencies mentioned above.  To this end, we consider two
conventional approaches used in this field: (i) the classical current
formalism, and (ii) symmetrization of the wave function.

(i) In the current formalism particle sources are described in terms
of a distribution of classical currents \cite{shuryak,podgor,gkw}.
For a Gaussian distribution of currents and in the absence of a
coherent component, all multiparticle distributions can be expressed
in terms of the two-current correlator, $\langle J^\star(x) J(x')
\rangle$ \cite{apw}.  The one- and two-particle inclusive
distributions then take the form\footnote{For neutral particles, there
may be an additional contribution to $P_2(\vec{k}_1,\vec{k}_2)$ which
only plays a role for soft particles and which will be neglected here,
cf. footnote 2 on page 4.}
\begin{equation}
P_1(\vec{k}) \ =\ \int d^4x \ w(x,k)
\label{eq:P1}
\end{equation}
and
\begin{equation}
P_2(\vec{k}_1,\vec{k}_2) \ =\ \int d^4x_1 \int d^4x_2
\left[w\left(x_1,k_1\right)
w\left(x_2,k_2\right)\ +\ w\left(x_1,K\right) w\left(x_2,K\right)
\ \exp\left[iq_\mu(x^\mu_1-x^\mu_2)\right] \right]
\label{eq:P2}
\end{equation}
where $K^\mu=(k_1^\mu+k_2^\mu)/2$ and $q^\mu=k_1^\mu-k_2^\mu$ are the
mean momentum and momentum difference of the pair, and where
\begin{equation}
w(x,k)
\ =\ \int d^4y \
\left\langle J^\star(x+\frac{y}{2}) J(x-\frac{y}{2}) \right\rangle
\ \exp\left[-ik_\mu y^\mu\right]
\end{equation}

(ii) In the wave function approach, the amplitude for emission of two
identical particles of momenta $k_1$, $k_2$ {}from the space-time
points $x_1$, $x_2$ is
\begin{eqnarray}
A(k_1,k_2) & = & M(k_1,x_1)M(k_2,x_2)
\exp[i(k_1x_1+k_2x_2)]\nonumber\\ & + & M(k_1,x_2)M(k_2,x_1)
\exp[i(k_1x_2+k_2x_1)]
\end{eqnarray}
The one- and two-particle inclusive distributions are
\begin{equation}
P_1(\vec{k}) \ =\ \int d^4x \ |M(k,x)|^2
\label{eq:P11}
\end{equation}
and
\begin{eqnarray}
P_2(\vec{k}_1,\vec{k}_2) & \equiv & \frac{1}{2 }\int d^4x_1 \int
d^4x_2 \
\left|A(k_1,k_2)\right|^2 \nonumber\\
& = & \int d^4x_1 \int d^4x_2 \quad
\left[\left|M(k_1,x_1)\right|^2 \left|M(k_2,x_2)\right|^2\
\right.\nonumber\\
& & + \ \left.  M(k_1,x_1) M^\star(k_2,x_1)M(k_2,x_2) M^\star(k_1,x_2)
\ \exp\left[iq_\mu(x^\mu_1-x^\mu_2)\right] \right]
\label{eq:P22}
\end{eqnarray}

Both expressions
\footnote{
To obtain specific results for $P_2(\vec{k}_1,\vec{k}_2)$ for photons
emitted {}from an expanding quark-gluon-plasma as discussed in
\cite{kapusta}, it is reasonable to substitute the emission
rates\cite{emi} for the integrands in Eqs. (\ref{eq:P1}) and
(\ref{eq:P11}), i.e., $$w(x,k)=|M(k,x)|^2= const. \ T^2 \ln
\left(\frac{2.9\ k_\mu u^\mu}{g^2T}
\ + \ 1\right) \ \exp\left[- \frac{k_\mu u^\mu}{T}\right]$$
where $u^\mu$ is the four velocity of the fluid, $T$ the temperature
and $g$ the QCD coupling constant.} (\ref{eq:P2}) and (\ref{eq:P22})
can be cast in the form
\begin{equation}
P_2(\vec{k}_1,\vec{k}_2)\ =\ P_1(\vec{k}_1) P_1(\vec{k}_2) \ +\
D(k_1,k_2) D^\star(k_1,k_2)
\label{eq:b}
\end{equation}
where for the current formalism
\begin{equation}
D(k_1,k_2) \ =\ \int d^4x \ w\left(x, \frac{k_1+k_2}{2}\right) \
\exp[- ix_\mu (k_1^\mu -k_2^\mu)]
\label{eq:di}
\end{equation}
and for the wave function approach
\begin{equation}
D(k_1,k_2) \ =\ \int d^4x \ M^\star(k_1,x) M(k_2,x) \ \exp[-ix_\mu
(k_1^\mu -k_2^\mu)]
\label{eq:dii}
\end{equation}
It then follows {}from Eq. (\ref{eq:b}) that $C_2(\vec{k}_1,\vec{k}_2)
\geq 1$ both for the results of (i) and of (ii).

The above considerations concerning bounds for the BEC functions refer
to the case of a Gaussian density matrix. In general, a different form
of the density matrix may yield correlation functions that are not
constrained by the bounds derived here. For instance, for a system of
squeezed states $C_2$ can take arbitrary positive values \cite{vour}.
Moreover, for particles produced in high energy hadronic or nuclear
collisions, the fluctuations of quantities such as impact parameter or
inelasticity may introduce additional correlations which may destroy
the Gaussian form of the density matrix and also affect the bounds of
the BEC functions (cf. ref. \cite{gyu}, \cite{LRC}).

\vspace*{1cm}

This work was supported by the Federal Minister of Research and
Technology under contract 06MR731, the Deutsche Forschungsgemeinschaft
and Gesellschaft f\"ur Schwer\-ionen\-for\-schung.


\end{document}